\documentclass[conference]{IEEEtran}
\IEEEoverridecommandlockouts
\pdfoutput=1
\usepackage{cite}
\usepackage{amsmath,amssymb,amsfonts}
\usepackage{algorithmic}
\usepackage{graphicx}
\usepackage{textcomp}
\usepackage{xcolor}
\usepackage{hyperref}
\usepackage{booktabs}
\usepackage{fancyhdr}
\usepackage[numbers]{natbib}
\renewcommand\harvardurl[1]{\textbf{URL:} \url{#1}}
\usepackage{hyperref}

\usepackage{subfiles}
\usepackage{tikz}
\usepackage{standalone}
\usetikzlibrary{fit}
\usetikzlibrary{positioning}
\usetikzlibrary{quotes, arrows.meta}

\def\BibTeX{{\rm B\kern-.05em{\sc i\kern-.025em b}\kern-.08em
    T\kern-.1667em\lower.7ex\hbox{E}\kern-.125emX}}

\begin{document}

\title{Equivariant Graph Attention Networks with Structural Motifs for Predicting Cell Line-Specific Synergistic Drug Combinations \\
\thanks{© 2024 IEEE.  Personal use of this material is permitted.  Permission from IEEE must be obtained for all other uses, in any current or future media, including reprinting/republishing this material for advertising or promotional purposes, creating new collective works, for resale or redistribution to servers or lists, or reuse of any copyrighted component of this work in other works.}
}

\author{\IEEEauthorblockN{Zachary Schwehr}
\IEEEauthorblockA{\textit{Mills E. Godwin}\\
Henrico, United States \\
zschwehr1@gmail.com}
}

\maketitle

\thispagestyle{plain}
\fancypagestyle{plain}{
\fancyhf{} 
\fancyfoot[L]{979-8-3503-5663-2/24/\$31.00~\copyright2024~IEEE} 
\renewcommand{\headrulewidth}{0pt}
\renewcommand{\footrulewidth}{0pt}
}

\begin{abstract}
Cancer is the second leading cause of death, with chemotherapy as one of the primary forms of treatment. As a result, researchers are turning to drug combination therapy to decrease drug resistance and increase efficacy. Current methods of drug combination screening, such as in vivo and in vitro, are inefficient due to stark time and monetary costs. In silico methods have become increasingly important for screening drugs, but current methods are inaccurate and generalize poorly to unseen anticancer drugs. In this paper, I employ a geometric deep-learning model utilizing a graph attention network that is equivariant to 3D rotations, translations, and reflections with structural motifs. Additionally, the gene expression of cancer cell lines is utilized to classify synergistic drug combinations specific to each cell line. I compared the proposed geometric deep learning framework to current state-of-the-art (SOTA) methods, and the proposed model architecture achieved greater performance on all 12 benchmark tasks performed on the DrugComb dataset. Specifically, the proposed framework outperformed other SOTA methods by an accuracy difference greater than 28\%. Based on these results, I believe that the equivariant graph attention network's capability of learning geometric data accounts for the large performance improvements. The model's ability to generalize to foreign drugs is thought to be due to the structural motifs providing a better representation of the molecule. Overall, I believe that the proposed equivariant geometric deep learning framework serves as an effective tool for virtually screening anticancer drug combinations for further validation in a wet lab environment. The code for this work is made available online at: \url{https://github.com/WeToTheMoon/EGAT_DrugSynergy}.
\end{abstract}

\begin{IEEEkeywords}
Graph Neural Networks, Attention, Equivariance, Structural Motifs, Combined Chemotherapy, Contrastive Learning
\end{IEEEkeywords}

\section{Introduction}
Cancer is the second leading cause of death and a massive barrier to increasing life expectancy \cite{ref1}. Current treatments fail to completely treat the disease due to adverse side effects and drug resistance. A primary treatment for cancer is the use of anticancer drugs to remove malignant cells through apoptosis and cellular death. However, these cancer cells develop escape methods and additional pathways for cell proliferation. As a result, scientists are looking to the use of multiple agents to treat different forms of cancer. The use of multiple drugs can overcome drug resistance through synergistic effects while decreasing toxicity and increasing efficacy \cite{ref2}. For instance, triple-negative breast cancer is a malignant type of cancer that has a high metastasis rate and poor prognosis. Lapatinib and Rapamycin are two different anticancer drugs that on their own have little effect when treating triple-negative breast cancer, however, can immensely increase the apoptosis rate of triple-negative breast cancer when used in tandem \cite{ref3}. On the contrary, other combinations of anticancer drugs are antagonistic and can even worsen the disease \cite{ref4}. On a biological level, chemotherapy drugs often work well together as they target different aspects or stages within cell division. The precise biological mechanisms that impact drug synergy are not well known, making it difficult to find synergistic drug combinations.

Current methods of discovering synergistic and antagonistic drug combinations are primarily based on experimental tests. These studies are time-consuming and costly, resulting in few drug combinations being screened. To fix these issues, high-throughput drug screening technology (HTS) allows researchers to simultaneously screen different drug combinations \cite{ref5}. However, results from HTS in vitro experiments heavily rely on the analysis with bioinformatics programs, preventing an accurate depiction of the drug’s mode of action in vivo \cite{ref6}. This caused researchers to turn to in silico methods. However, current in silico methods yield poor accuracy and do not model drug interactions well.
	
The rise of large datasets allows for the production of in silico models to predict synergistic combinations of anti-cancer drugs. These models tend to utilize the genetic information of the cells as well as the chemical properties of the different drugs. Complex algorithms, such as deep learning frameworks, have been shown to have an increased performance. For example, DeepSynergy uses a feed-forward neural network to combine the gene expression data from the cancer cell line and the molecular representations of each drug \cite{ref8}.

Furthermore, AuDNNsynergy employs three autoencoders for the mutation, gene expression, and copy number variation data \cite{ref9}. Graph Neural Networks (GNNs), have also been applied to predict synergy such as DeepDDS which uses attention mechanisms with GNNs \cite{ref21}. In these graphs, the atoms act as the nodes, and the bonds between the atoms represent the edges. GNNs have been used in other molecular tasks, such as predicting toxicity and binding affinity due to their ability to learn molecular features. Others have also employed geometric transformer architectures which obtain edge information typically based on geometric data such as in tasks involving proteins.

I propose an equivariant GNN with attention mechanisms and structural motifs. The proposed model is trained on binary labels, including samples of the DrugComb dataset. The DrugComb dataset is one of the largest synergy datasets which allows the model to learn and be tested on a wide variety of data \cite{ref10}. With the dataset, the model is trained using supervised contrastive learning followed by binary cross-entropy. Unlike previous frameworks, the proposed model computes its own representation of each drug using message passaging schemes instead of predetermined chemical features. An additional algorithm is used to find structural motifs. These structural motifs represent the chemical groups of the drug, allowing for a greater generalizability of larger molecules. The model outperforms various state-of-the-art baseline models when tested on benchmark datasets.

\section{Methods and Materials}

\subsection{Dataset}
The most comprehensive benchmark dataset for predicting synergistic drug combinations is the DrugComb dataset \cite{ref11}. The DrugComb dataset is a web-based portal containing the analysis and information on various drug combination screening datasets. In total, the dataset contains combinations from over $8000$ drugs and $2320$ cancer cell lines. The objective of the DrugComb dataset is to predict synergistic and antagonistic drug combinations given the SMILES string and the cancer cell line \cite{ref12}. The gene expression for each cell line was obtained from the Cancer Cell Line Encyclopedia, an independent dataset containing normalized mRNA expression data \cite{ref13}. 

\subsection{Loewe Additivity Model}
Synergy scores are calculated based on the response percent beyond the calculated expected values. The method of calculating these expected values varies with the different synergy scores. One synergy score, the Loewe additivity model (LAM), is built on the concepts of sham combination and dose equivalence. The sham combination states that the compound cannot interact with itself, while dose equivalence contends that the same effect of both compounds is exchangeable. Based on LAM, the Loewe additive response is calculated as:
\begin{equation}
Y_{\text{LAM}}=\frac{P_{\text{min}}+P_{\text{max}} \left(\frac{d_{1}+d_{2}}{m}\right)^{\lambda}}{1+\left(\frac{d_{1}+d_{2}}{m}\right)^\lambda}
\end{equation}
\noindent where $Y_{\text{LAM}}$ is the loewe additivity response, $P_{\text{min}}$ and $P_{\text{max}}$ are the minimum and maximum pharmacodynamics response, respectively, and $d_{1}+d_{2}$ are the doses of drugs 1 and 2. $\lambda$ is the shape parameter and $m$ is the dosage that would produce the midpoint response between $P_{\text{min}}$ and $P_{\text{max}}.$ 

\subsection{Bliss Independence Model}
The Bliss Independence Model (BIM) is employed as an alternative to LAM. The primary concept of BIM is that it assumes that the two drugs do not produce an equal effect in treating the disease. The drug response has a direct correlation to the amount of the drug. Therefore, the bliss response of the drug combination can be computed as:
\begin{equation}
Y_{\text{BIM}}= P_{1} + P_{2} - P_{1}P_2
\end{equation}
\noindent where $P_{1}$ and $P_{2}$ are the pharmacodynamics responses of drugs 1 and 2, respectively.

\subsection{Highest Single Agent Model}
The highest single agent model states that the combined drug response is equal to the greatest drug response of the drugs. The highest single agent is calculated as:
\begin{equation}
Y_{\text{HSAM}} = \text{max}\left(P_{1},P_{2}\right)
\end{equation}
\noindent where all the variables are defined in BIM. 

\subsection{Zero Interaction Potency Model}
The zero interaction potency model (ZIPM) utilizes the concepts within LAM and BIM through logistic functions as:
\begin{equation}
Y_{\text{ZIPM}}=\frac{\left(\frac{d_{1}}{m_1}\right)^{\lambda_1}}{1+\left(\frac{d_{1}}{m_1}\right)^{\lambda_1}}+\frac{\left(\frac{d_{2}}{m_2}\right)^{\lambda_2}}{1+\left(\frac{d_{2}}{m_2}\right)^{\lambda_2}}-\frac{\left(\frac{d_{1}}{m_1}\right)^{\lambda_1}}{1+\left(\frac{d_{1}}{m_1}\right)^{\lambda_1}}\frac{\left(\frac{d_{2}}{m_2}\right)^{\lambda_2}}{1+\left(\frac{d_{2}}{m_2}\right)^{\lambda_2}}
\end{equation}
\noindent where all the variables are defined in LAM \cite{ref10}. The presence of a high and low synergy score for the four different models, Zip, Bliss, HSA, and Loewe, would indicate a synergistic and antagonistic relationship between the chemotherapy drugs, respectively.

\subsection{Drug Representations}
In the DrugComb dataset, the drugs were represented as SMILES strings \cite{ref12}. RDKit was used to convert the SMILES strings into molecular graphs where the nodes are the vertices and the bonds are the edges \cite{ref14}. Drugs were represented as graphs defined as $\mathcal{G} = (\mathcal{V},\mathcal{E})$, where $\mathcal{V}$ is the set of nodes, $\mathcal{N}$, which are represented by a d-dimensional vector. $\mathcal{E}$ is the set of edges represented as an adjacency matrix $A$ and edge attributes $a_{ij}$. In the molecular graph, $n_i \in \mathcal{V}$ represents the $i$-th atom. The chemical bond and chemical bond attributes between the $i$-th and $j$-th atom are denoted as $e_{ij} \in \mathcal{E}$ and $a_{ij}$, respectively. Furthermore, each atom, $n_i$, also has a corresponding 3D coordinate, $x_i$ which was also calculated using RDKit. 

Each atom, $n_i$, is represented using a feature vector, $h_{i}^{l} \in \mathbb{R}^d$, containing information about that atom: atomic symbol, electronegativity, atomic radius, hybridization, degree, formal charge, number of radical electrons, number of hydrogens, chirality, chirality type, and aromaticity. The atomic symbol, hybridization, degree, number of hydrogens, chirality, chirality type, and aromaticity were represented as one-hot encoded vectors. Each edge attribute, $a_{ij}$, was represented using the bond type, aromaticity, conjugation, and whether it was in a ring. The model did not represent all of the atoms, only those that were in the training data. An additional atomic symbol was represented as "other" for the atoms that were not present in the training data. The drugs were further encoded using a graph neural network as in Fig. 1.

\subsection{Graph Neural Network}
Similar to feed-forward networks, GNNs contain multiple layers $L$, signifying the depth of the neural network. Each layer, $l \in \left\{1, ..., L\right\}$, specifies that each node, $n_i$ can only obtain information from $l$ nodes away. Neighboring nodes are denoted as $\mathcal{N}(i)$ where each node vector representation, $h_{i}^{l-1}$, is updated at layer $l$ through the aggregation of the neighboring messages:

\begin{equation}
\begin{aligned}
& m_{ij}^{l} = \phi^{l}\left(h_{i}^{l-1}, h_{j}^{l-1}, a_{ij}\right) \\
& m_{i}^l = \sum_{j \in \mathcal{N}\left(i\right)}^{}m_{ij}^{l} \\
& h_{i}^{l} = \gamma^{l}\left(h_{i}^{l-1}, m_{i}^{l}\right)\\
\end{aligned}
\end{equation}

\noindent where $m_{ij}^l$ represents the message from $n_{j}^{l-1}$ to  $n_{i}^{l-1}$. The aggregation function is a permutation invariant function that aggregates all the messages, $m_{ij}^l$, with one of the most common aggregation functions: summation. The messages, $m_{ij}^{l}$ are calculated using $\phi^{l}$, and $h_{i}^{l-1}$ is updated using $\gamma^{l}$ which represents a multi-layer perception (MLP) \cite{ref15}. 

\subsection{Graph Attention Network}
The graph attention network (GAT) utilizes a multi-head attention-based architecture that attempts to learn higher-level features of the different nodes through the use of self-attention mechanism. The graph attention layer computes attention coefficients which weigh the importance of the connection between the $i$-th and $j$-th node. These single-head attention coefficients are calculated as such:

\begin{equation}
\begin{aligned}
& e\left(h_{i}^{l-1}, h_{j}^{l-1}\right) = \text{LeakyReLU} \left( \overrightarrow{a}^\top\cdot \left[ Wh_{i}^{l-1}||Wh_{j}^{l-1}\right] \right) \\
& \alpha_{ij}^l = \frac{\text{exp}\left(e\left(h_{i}^{l-1}, h_{j}^{l-1}\right)\right)}{\sum_{j'\in \mathcal{N}_i}^{}\text{exp}\left(e\left(h_{i}^{l-1}, h_{j'}^{l-1}\right)\right)}
\end{aligned}
\end{equation}

\noindent where $\overrightarrow{a} \in \mathbb{R}^{2d'}$ and $W \in \mathbb{R}^{d \times d'}$ are learned and $||$ is the vector concatenation operation \cite{ref16}. These attention coefficients are then used during aggregation as in:

\begin{equation}
m_{i}^{l} = \sum_{j \in \mathcal{N}_i}^{}\alpha_{ij}^l \space \cdot \space m_{ij}^l
\end{equation}

Brody et al. have proposed the GATv2 which computes dynamic attention coefficients increasing the GAT's expressiveness. The original GAT applies a linear transformation of $W$ prior to concatenation, which is then followed by the linear transformation with $\overrightarrow{a}$. This process is the same as applying these linear transformations, $W$ and $\overrightarrow{a}$, consecutively, which can be performed in a single linear transformation. This leads to static attention coefficients as one key tends to have greater attention coefficients for all of their queries. 

\quad In GATv2, the linear transformation with $\overrightarrow{a}$ is performed following the nonlinearity (LeakyReLU). This sequence of operations allows the model to learn attention coefficients  effectively for each query-key pair using an MLP instead of a single linear transformation. The use of an MLP instead of a linear transformation allows for dynamic attention coefficients instead of static ones. The GATv2 layer is expressed as:

\begin{equation}
e\left(h_{i}^{l-1}, h_{j}^{l-1}\right) = \overrightarrow{a}^\top \cdot \text{LeakyReLU}\left(W \cdot \left[h_{i}^{l-1}||h_{j}^{l-1}\right]\right)
\end{equation}

\noindent where all variables are the same as those in GAT. In this experiment, the GATv2 is extended to multi-head attention to stabilize training and improve generalizability: 

\begin{equation}
\mathrm{\overset{K}{\underset{k=1}{\Big|\Big|}}\sigma \left(\sum_{j \in \mathcal{N}_i}^{}\alpha_{ij}^{lk} \space \cdot \space m_{ij}^{lk}\right)
}
\end{equation}

\noindent where $K$ is the number of heads and $\sigma$ is a nonlinearity function. In this equation, the multi-head attention coefficients are concatenated, however, these heads can be summated or aggregated in different occasions. 

\begin{figure*}
    \begin{center}
        \input{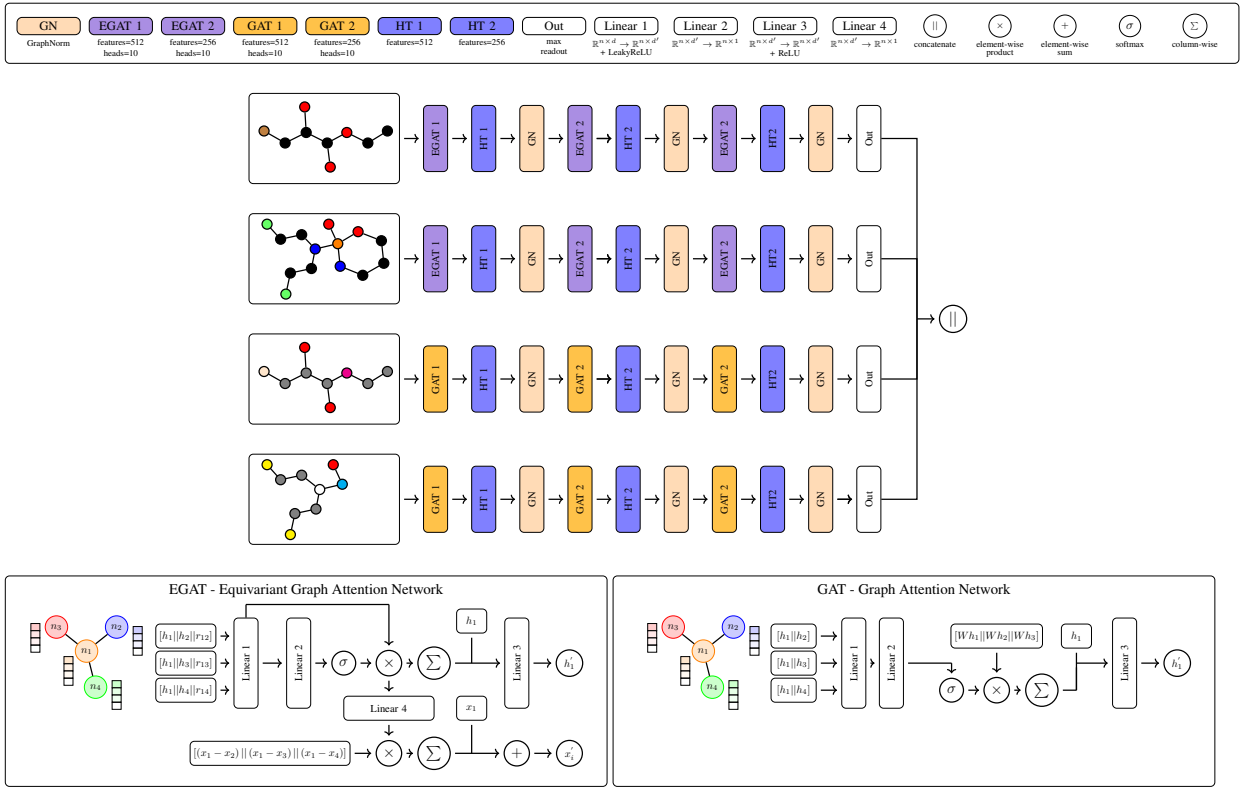}
        \caption{The framework for encoding the two drugs. The two drugs and their two motif structures are each encoded using the EGAT and GAT, respectively. Following each graph layer, the attention heads are aggregated using a linear transformation with the same number of feature channels, followed by a graph normalization layer. The EGAT and GAT layers have similar structures, however, the EGAT updates the coordinate values while maintaining equivariance. Following the graph layers, the graph-level features are expressed using maximum readout and then concatenated.}
    \end{center}
\end{figure*}

\subsection{Equivariance}
Given transformations $T_g : \mathcal{X} \to \mathcal{X}$ for the abstract group $g \in G$, a function $\phi : \mathcal{X} \to \mathcal{Y}$ is equivariant for all g if there exists a transformation $S_g : \mathcal{Y} \to \mathcal{Y}$ such that:

\begin{equation}
\phi\left(T_g\left(x\right)\right) = S_g\left(\phi\left(x\right)\right) \indent \forall g \in G, \forall x \in \mathcal{X}
\end{equation}

\noindent Invariance is similar to equivariance, where the transformation does not affect the prediction such that:

\begin{equation}
\phi\left(T_g\left(x\right)\right) = \phi\left(x\right) \indent \forall g \in G, \forall x \in \mathcal{X}
\end{equation}

In this literature, I employ Satorras et al's Equivariant Graph Neural Network (EGNN) which is E(n) equivariant: translation, rotation, and permutation equivariant. Assuming a graph with $N$ nodes each with a coordinate $x_i \in \mathbb{R}^n$, translation equivariance is defined as $y + g = \phi (x+g)$ where $g \in \mathbb{R}^n$ and $y \in \mathbb{R}^{N \times n}$. Rotation and reflection equivariance is defined as $Qy = \phi (Qx)$ where $Q \in \mathbb{R}^{n \times n}$ is an orthogonal matrix. Permutation equivariance is defined as $P(y) = \phi (P(X))$ where P permutates the row indexes \cite{ref17}.

\subsection{Equivariant Graph Attention Network}

Satorras et al's EGNN employs a message passaging system similar to that of a graph convolution network, but it incorporates geometric and positional information during message passaging. It utilizes node features $h_i^{l-1}$, node-coordinates $x_i^{l-1}$, edges $e_{ij}$, and edge attributes $a_{ij}$. The EGNN's message passaging framework is as such:

\begin{equation}
\begin{aligned}
& m_{ij}^{l} = \phi^{l}\left(h_{i}^{l-1}, h_{j}^{l-1}, \left\| x_i^{l-1}-x_j^{l-1} \right\|_2^2, a_{ij}\right) \\
& x_i^{l} = x_i^{l-1} + \sum_{j\neq i}^{}\left(x_i^{l-1}-x_j^{l-1}\right) \space \varphi^l\left(m_{ij}^l\right)\\
& m_{i}^l = \sum_{j \in \mathcal{N}\left(i\right)}^{}m_{ij}^{l} \\
& h_{i}^{l} = \gamma^{l}\left(h_{i}^{l-1}, m_{i}^{l}\right)\\
\end{aligned}
\end{equation}

\noindent where $\phi^{l}$, $\varphi^l$, and $\gamma^{l}$ are MLPs. With this message passaging scheme, E(n) equivariance remains \cite{ref17}.

In this work, I merged the equivariant message passaging scheme with dynamic multi-headed attention coefficients, as in the reworked graph attention network (GATv2), to create an E(N) Equivariant Graph Attention Network (EGAT). The EGAT is equivariant to 3D rotations, translations, and reflections. The computation of the attention coefficients for the EGAT are the same as those in (6) and the message passaging scheme with single-head self-attention mechanisms for the EGAT is as such:

\begin{equation}
\begin{aligned}
& e(h_{i}^{l-1}, h_{j}^{h-1}, r_{ij}^{l-1}, a_{ij}) = \vec{a}^\top \cdot \sigma \left(W \cdot \left[h_{i}^{l-1}||h_{j}^{l-1}|| r_{ij}^{l-1}|| a_{ij} \right]\right) \\
& \alpha_{ij}^l = \frac{\text{exp}(e(h_{i}^{l-1}, h_{j}^{l-1}, r_{ij}^{l-1}, a_{ij}))}{\sum_{j'\in \mathcal{N}_i}^{}\text{exp}(e(h_{i}^{l-1}, h_{j}^{l-1}, r_{ij}^{l-1}, a_{ij}))} \\
& m_{ij}^{l} = \phi^{l}(h_{i}^{l-1}, h_{j}^{l-1}, r_{ij}^{l-1}, a_{ij}) \\
& x_i^{l} = x_i^{l-1} + \sum_{j\neq i}^{}(x_i^{l-1}-x_j^{l-1}) \space \varphi^l(\alpha_{ij}^l \cdot m_{ij}^l)\\
& m_{i}^l = \sum_{j \in \mathcal{N}(i)}^{}\alpha_{ij}^l \cdot m_{ij}^{l} \\
& h_{i}^{l} = \gamma^{l}(h_{i}^{l-1}, m_{i}^{l}) \\
\end{aligned}
\end{equation}

\noindent where $r_{ij}^{l-1} = \left\| x_i^{l-1}-x_j^{l-1} \right\|_2^2$, $\sigma$ is the LeakyReLU nonlinearity function, and all other variables are the same as in (12). When expanding the EGAT to multi-head attention, only one head is used when updating the positional coordinate, $x_i^{l-1}$, which is a vector field in a radial direction.

\subsection{Graph Normalization}
In the proposed framework, I implement Graph Normalization, proposed by Cai et al which proposed an alternate normalization method for graphs. They showed that Graph Normalization converges faster compared to other common normalization methods: BatchNorm and InstanceNorm. This was believed to be the case due to the heavy batch noise in BatchNorm and the degradation of expressiveness found in InstanceNorm for regular graphs. Graph Normalization is as such:

\begin{equation}
    GraphNorm(\hat{h}_{ik}) = \zeta_k \cdot \frac{\hat{h}_{ik}-\psi_k \cdot \mu_k}{\sigma_k} - \beta_k
\end{equation}

\noindent where $\hat{h}_{ik}$ is the input which denotes the $k$-th feature value of the $i$-th node, $\mu_k=\frac{\sum^{n}_{i=1}\hat{h}_{ik}}{n}$, $\sigma_k=\frac{\sum^{n}_{i=1}\left( \hat{h}_{ik} - \psi_k \cdot \mu_k \right)^2}{n}$, and $\zeta_k$, $\beta_k$, and $\psi_k$ are learnable parameters. $\zeta_k$ and $\beta_k$ are affine parameters that are also present in BatchNorm and InstanceNorm, and $\psi_k$ represents the amount of information needed to be kept in the mean for each feature dimension $k$ \cite{ref18}.

\subsection{Structural Motifs}
Organic compounds and drugs are typically made of smaller building blocks: functional groups. As such, many drugs share similar functional groups and rings. To extract these common functional groups and patterns, I implemented structural motifs to extract more information and increase the model's generalizability. 

\quad Similar to Jin et al, I define a motif $\mathcal{S}_i=(\mathcal{V}_i, \mathcal{E}_i)$ as a subgraph of the molecule $\mathcal{G}$ \cite{ref23}. Given a molecule $\mathcal{G}$, structural motifs $\mathcal{S}_1, \cdots, \mathcal{S}_n $ are extracted such that the collection of motifs fully represents $\mathcal{G}$. These motifs, $\mathcal{S}_i$, contain rings and elements that are not within another ring. Based on these rules, a motif dictionary is extracted and the motifs with a frequency less than 100 were removed. An additional motif, "other," was implemented for atoms that were not present in the training data. Using the motif dictionary, the molecules $\mathcal{G}$ were decomposed such that the motif representation comprised of subgraphs $\mathcal{S}_i$ fully representing the molecule $\mathcal{G}$.

\subsection{Supervised Contrastive Learning}
The most common loss function for binary classification tasks is binary cross-entropy. However, a different approach has been proposed: supervised contrastive learning. One of the greatest drawbacks to cross-entropy loss is the lack of robustness towards noisy labels which decreases generalizability and performance. Supervised contrastive learning has attempted to solve these shortcomings by pulling together the shared labels within the embedding space and pushing away the uncommon labels \cite{ref19}. 

\quad The InfoMCE loss function pushes and pulls these samples within the embedding space. Given an encoded query $q$ and a set of encoded samples $\left\{ s_0, s_1, s_2, ... s_i\right\}$, there is a sample $s_+$ that matches $q$. The InfoMCE loss function determines the similarity between $q$ and $s_+$ and the dissimilarity between $q$ and all other samples. The InfoMCE is such as:

\begin{equation}
\mathcal{L}_E = -\text{log}\left( \frac{\text{exp}\left(q \cdot s_+ / \tau \right)}{\sum_{i=0}^{N}\text{exp}\left(q \cdot s_i / \tau \right)} \right)
\end{equation}

\noindent where similarity is determined using dot product, $\tau$ is a hyperparameter, and $N$ is the number of samples that are not $s_+$ \cite{ref20}. In this study, $\tau$ was set to $0.1$. Once the encoder is trained based on the contrastive loss function, binary cross entropy is implemented to train the classifier as such:

\begin{equation}
\mathcal{L}_C = -y_i\text{log}\left( \hat{y_i} \right) + \left( 1-y_i \right) \text{log} \left( 1-\hat{y_i} \right)
\end{equation}

\noindent where $\hat{y}_i$ is the predicted probability for the i-th sample and $y_i \in [0, 1]$ is the true label. Due to the clustering of the embedding space, the training of the classifier is substantially easier.

\begin{table}
    \begin{center}
    \caption{Hyperparameters}
    \begin{tabular}{l l}
        \toprule
        Hyperparameter & Value \\
        \midrule
        \# EGAT and GAT Layers & 3 \\
        EGAT Encoder & $\left[ 512, 256, 256 \right]$ \\
        GAT Encoder & $\left[ 64, 32, 32 \right]$ \\
        Dropout & 0.2 \\
        Heads & $6$ \\
        Cell Line Encoder & $\left[ 2048, 1024, 576 \right]$ \\
        Classification Head & $\left[ 576, 512, 256, 128 \right]$ \\
        \bottomrule
    \end{tabular}
    \end{center}
\end{table}

\subsection{Training}

\begin{table*}
\begin{center}
\caption{Comparison to SOTA - Loewe Synergy Score. Top-2 are in \textcolor{red}{RED} and \textcolor{blue}{BLUE}.}
\begin{tabular}{ l c c c c c c c}
\toprule
 & \multicolumn{3}{c}{Transductive} & \multicolumn{2}{c}{Unknown Combination} & \multicolumn{2}{c}{Unknown Drug} \\
\cmidrule(l{7.5em}r{0.3em}){1-8}
Method & AUROC & ACC & AUPRC & AUROC & ACC & AUROC & ACC \\
\midrule
Proposed & \textcolor{red}{$94.55 \pm 0.13$} & \textcolor{red}{$92.93 \pm 0.10$} & \textcolor{red}{$82.80 \pm 0.37$} & \textcolor{red}{$85.50 \pm 0.96$} & \textcolor{red}{$91.27 \pm 0.41$} & \textcolor{red}{$82.59 \pm 0.86$} & \textcolor{red}{$92.41 \pm 0.70$} \\
DeepDDS & $76.82 \pm 0.87$ & $84.98 \pm 0.11$ & \textcolor{blue}{$50.54 \pm 1.67$} & \textcolor{blue}{$74.96 \pm 1.00$} & \textcolor{blue}{$84.97 \pm 0.10$} & \textcolor{blue}{$77.47 \pm 1.23$} & \textcolor{blue}{$85.13 \pm 0.66$} \\
DeepSynergy & \textcolor{blue}{$79.39 \pm 0.75$} & \textcolor{blue}{$86.22 \pm 0.33$} & $49.87 \pm 2.07$ & $70.66 \pm 1.21$ & \textcolor{blue}{$84.97 \pm 0.10$} & $73.87 \pm 1.53$ & $83.03 \pm 0.84$ \\ 
XGBoost & $49.35 \pm 0.28$ & $85.45 \pm 0.51$ & $52.47 \pm 2.23$ & $49.45 \pm 0.24$ & $84.82 \pm 0.32$ & $48.88 \pm 0.36$ & $84.78 \pm 0.15$ \\
LogReg & $50.73 \pm 1.45$ & $75.03 \pm 7.89$ & $49.84 \pm 1.94$ & $52.26 \pm 1.80$ & $77.34 \pm 4.66$ & $52.15 \pm 2.04$ & $77.86 \pm 4.81$ \\

\bottomrule
\end{tabular}
\end{center}
\end{table*}

\begin{table*}
\begin{center}
\caption{Comparison to SOTA - HSA Synergy Score. Top-2 are in \textcolor{red}{RED} and \textcolor{blue}{BLUE}.}
\begin{tabular}{ l c c c c c c c}
\toprule
 & \multicolumn{3}{c}{Transductive} & \multicolumn{2}{c}{Unknown Combination} & \multicolumn{2}{c}{Unknown Drug} \\
\cmidrule(l{7.5em}r{0.3em}){1-8}
Method & AUROC & ACC & AUPRC & AUROC & ACC & AUROC & ACC \\
\midrule
Proposed & \textcolor{red}{$97.45 \pm 0.22$} & \textcolor{red}{$93.32 \pm 0.10$} & \textcolor{red}{$95.21 \pm 0.29$} & \textcolor{red}{$92.18 \pm 0.56$} & \textcolor{red}{$91.79 \pm 0.35$} & \textcolor{red}{$88.97 \pm 0.83$} & \textcolor{red}{$88.05 \pm 0.70$}\\
DeepDDS & \textcolor{blue}{$96.10 \pm 0.27$} & $68.73 \pm 0.24$ & \textcolor{blue}{$93.15 \pm 0.28$} & \textcolor{blue}{$71.23 \pm 0.62$} & $68.28 \pm 0.34$ & \textcolor{blue}{$70.76 \pm 0.87$} & \textcolor{blue}{$68.20 \pm 1.90$} \\
DeepSynergy & $84.76 \pm 0.08$ & \textcolor{blue}{$81.87 \pm 0.08$} & $77.76 \pm 0.08$ & $52.16 \pm 1.78$ & \textcolor{blue}{$68.40 \pm 0.36$} & $51.21 \pm 0.54$ & $66.34 \pm 1.34$ \\ 
XGBoost & $49.71 \pm 0.12$ & $63.85 \pm 3.03$ & $53.21 \pm 0.18$ & $49.17 \pm 0.42$ & $68.31 \pm 0.41$ & $49.49 \pm 0.44$ & $63.03 \pm 4.75$ \\
LogReg & $49.64 \pm 1.33$ & $55.14 \pm 1.07$ & $51.21 \pm 0.92$ & $51.87 \pm 2.05$ & $59.92 \pm 2.78$ & $51.11 \pm 1.49$ & $58.89 \pm 2.58$ \\
\bottomrule
\end{tabular}
\end{center}
\end{table*}

\begin{table*}
\begin{center}
\caption{Comparison to SOTA - ZIP Synergy Score. Top-2 are in \textcolor{red}{RED} and \textcolor{blue}{BLUE}.}
\begin{tabular}{ l c c c c c c c}
\toprule
 & \multicolumn{3}{c}{Transductive} & \multicolumn{2}{c}{Unknown Combination} & \multicolumn{2}{c}{Unknown Drug} \\
\cmidrule(l{7.5em}r{0.3em}){1-8}
Method & AUROC & ACC & AUPRC & AUROC & ACC & AUROC & ACC \\
\midrule
Proposed & \textcolor{red}{$98.58 \pm 0.03$} & \textcolor{red}{$94.86 \pm 0.06$} & \textcolor{red}{$99.01 \pm 0.03$} & \textcolor{red}{$96.94 \pm 0.21$} & \textcolor{red}{$94.05 \pm 0.13$} & \textcolor{red}{$92.67 \pm 0.87$} & \textcolor{red}{$89.45 \pm 0.52$} \\
DeepDDS & \textcolor{blue}{$97.60 \pm 0.13$} & $70.89 \pm 1.69$ & \textcolor{blue}{$98.37 \pm 0.14$} & \textcolor{blue}{$74.19 \pm 0.54$} & \textcolor{blue}{$60.31 \pm 0.58$} & \textcolor{blue}{$67.97 \pm 0.51$} & \textcolor{blue}{$61.00 \pm 1.23$} \\
DeepSynergy & $80.20 \pm 0.23$ & \textcolor{blue}{$72.67 \pm 0.41$} & $84.91 \pm 0.22$ & $58.17 \pm 5.01$ & $60.25 \pm 0.56$ & $51.80 \pm 1.11$ & $59.7 \pm 1.06$ \\ 
XGBoost & $50.09 \pm 0.39$ & $60.54 \pm 0.32$ & $53.21 \pm 0.51$ & $49.68 \pm 0.21$ & $56.53 \pm 3.53$ & $49.78 \pm 0.22$ & $60.86 \pm 0.82$ \\
LogReg & $51.34 \pm 1.07$ & $56.00 \pm 2.34$ & $55.21 \pm 0.24$ & $50.48 \pm 1.00$ & $52.58 \pm 2.99$ & $51.63 \pm 1.27$ & $57.97 \pm 1.10$ \\
\bottomrule
\end{tabular}
\end{center}
\end{table*}

\begin{table*}
\begin{center}
\caption{Comparison to SOTA - Bliss Synergy Score. Top-2 are in \textcolor{red}{RED} and \textcolor{blue}{BLUE}.}
\begin{tabular}{ l c c c c c c c}
\toprule
 & \multicolumn{3}{c}{Transductive} & \multicolumn{2}{c}{Unknown Combination} & \multicolumn{2}{c}{Unknown Drug} \\
\cmidrule(l{7.5em}r{0.3em}){1-8}
Method & AUROC & ACC & AUPRC & AUROC & ACC & AUROC & ACC \\
\midrule
Proposed & \textcolor{red}{$98.07 \pm 0.06$} & \textcolor{red}{$93.48 \pm 0.15$} & \textcolor{red}{$97.95 \pm 0.06$} & \textcolor{red}{$93.24 \pm 1.45$} & \textcolor{red}{$88.95 \pm 1.78$} & \textcolor{red}{$89.90 \pm 1.23$} & \textcolor{red}{$84.92 \pm 1.13$} \\
DeepDDS & \textcolor{blue}{$96.53 \pm 0.17$} & \textcolor{blue}{$88.47 \pm 1.17$} & \textcolor{blue}{$96.49 \pm 0.19$} & \textcolor{blue}{$69.03 \pm 1.35$} & \textcolor{blue}{$65.37 \pm 1.56$} & \textcolor{blue}{$62.7 \pm 1.62$} & \textcolor{blue}{$59.94 \pm 1.59$} \\
DeepSynergy & $73.98 \pm 0.17$ & $66.41 \pm 1.05$ & $75.35 \pm 0.17$ & $50.24 \pm 0.24$ & $50.15 \pm 0.31$ & $50.97 \pm 0.53$ & $50.22 \pm 0.71$ \\ 
XGBoost & $50.86 \pm 0.86$ & $50.22 \pm 0.45$ & $51.26 \pm 0.91$ & $50.73 \pm 0.73$ & $50.14 \pm 0.55$ & $49.88 \pm 0.12$ & $50.82 \pm 0.33$ \\
LogReg & $52.19 \pm 1.30$ & $52.15 \pm 1.34$ & $54.26 \pm 0.68$ & $51.87 \pm 1.16$ & $51.03 \pm 1.50$ & $52.61 \pm 1.19$ & $52.28 \pm 1.28$ \\
\bottomrule
\end{tabular}
\end{center}
\end{table*}

The model was trained using the Adam optimizer, for 450 epochs with a batch size of 128 and a learning rate of $0.0001$. The hyperparameters for the model architecture can be seen in Table I. Experimentation results were achieved using an Intel Core I7 processor running at 3.6 GHz, 64 GB RAM, and an NVIDIA 3090 GPU running on a 64-bit operating system. The data was split using 5-fold cross-validation. The models were accessed using AUROC, accuracy, and AUPRC. 
\section{Results}

The model was tested on the DrugComb dataset and it outperformed other state-of-the-art (SOTA) models on all of the tested benchmarks, based on the AUROC and accuracy metrics. The AUPRC metric was also implemented on the transductive datasets due to the importance of precision and recall in medical diagnosis, treatment, and prognosis. The SOTA models that were compared to include DeepDDS, DeepSynergy, Logistic Regression, and XGBoost \cite{ref21, ref22}. For DeepSynergy, Logistic Regression, and XGBoost, three graph attention layers were implemented to extract graph-level features. The graph attention layers were trained with supervised contrastive learning, similar to that of the proposed model.

\quad The benchmarks include the four different synergy scores, ZIP, Loewe, HSA, and Bliss, as well as three separate dataset splits: transductive, unknown combination, and unknown drug. In the unknown combination dataset, the data was split such that each of the five folds were roughly equal and the training set excluded all drug combinations from the test set. The unknown drug dataset had the same format as the unknown combination dataset, but the test set included only the drugs that the model did not train on. Table II shows the quantitative results of the transductive, unknown combination, and unknown drug datasets for Loewe Synergy. Table III shows the quantitative results of the transductive, unknown combination, and unknown drug datasets for HSA Synergy. Table IV shows the quantitative results of the transductive, unknown combination, and unknown drug datasets for Zip Synergy. Table V shows the quantitative results of the transductive, unknown combination, and unknown drug datasets for Bliss Synergy. All four tables show the quantitative results including the mean of the 5-fold cross-validation as well as the standard error.

\quad The proposed model displays significant performance improvements compared to the other SOTA methods, most predominant in the unknown combination and drug datasets. Specifically, the model achieves the greatest accuracy and AUROC amongst all 12 datasets. In some datasets, the model (94.05\%) achieves an accuracy of up to 33\% greater than the second best (60.31\%). This increased performance on these datasets exhibits the model's increased expressivity and generalizability. 

\begin{table*}
\centering
\caption{Ablation Study on the Transductive Loewe Study}
\hspace*{-4em}
\begin{tabular}{c | c c c | c c }
\toprule
\multicolumn{1}{c|}{Model} & \multicolumn{3}{c|}{Ablation} & \multicolumn{2}{c}{Transductive Loewe} \\
 
Number & Equivariance & Structural Motifs & Attention & AUROC & ACC \\
\midrule
(1) & & & & $91.94 \pm 0.11$ & $90.36 \pm 0.09$ \\ 
(2) & $\checkmark$ & & & $92.83 \pm 0.13$ & $91.3 \pm 0.05$ \\ 
(3) &  & $\checkmark$ & & $92.91 \pm 0.10$ & $91.19 \pm 0.10$ \\ 
(4) &  & & $\checkmark$ & $93.11 \pm 0.10$ & $91.45 \pm 0.04$ \\ 
(5) & $\checkmark$ & $\checkmark$ & & $93.46 \pm 0.08$ & $92.05 \pm 0.12$ \\ 
(6) & $\checkmark$ & & $\checkmark$ & \textcolor{blue}{$94.47 \pm 0.04$} & \textcolor{blue}{$92.90 \pm 0.07$} \\ 
(7) &  & $\checkmark$ & $\checkmark$ & $94.06 \pm 0.24$ & $92.43 \pm 0.18$ \\ 
(8) & $\checkmark$ & $\checkmark$ & $\checkmark$ & \textcolor{red}{$94.55 \pm 0.13$} & \textcolor{red}{$92.93 \pm 0.10$} \\ 
\bottomrule
\end{tabular}
\end{table*}

\quad Furthermore, of the SOTA models, DeepDDS, a GNN-based architecture, has the second greatest ability to generalize due to its performance on the unknown combination and unknown drug datasets. The significant increase in performance compared to the SOTA models on the 12 benchmark datasets shows the superiority of the E(N) equivariance and structural motifs in the GNN. The structural motifs allow the model to better represent common functional groups and patterns within the chemotherapy drugs, allowing the model to learn stronger molecular features. Additionally, maintaining rotational and translation equivariance increases the model's robustness.

\quad Within the unknown combination and unknown drug datasets, the disparity between the proposed model and the second-best performing models increases compared to that of the transductive datasets. This shows that the other SOTA models fail to encode molecular relationships and molecules that have not been exposed to prior. This large disparity between the SOTA models and the proposed model shows the expressiveness of the structural motifs as well as the equivariant layers.

\subsection{Ablation Study}

\quad I ran multiple ablation studies in Table VI to evaluate the efficacy and performance improvements of each of the implemented methods in the proposed model including multi-headed self-attention, E(N) equivariance, and structural motifs. I compared the final proposed model to several other models that implemented the different methods in the ablation study. The baseline model (1) is a graph neural network with message passaging without attention mechanisms and structural motifs, and it is not equivariant. Models (2)-(4) contain only one of the implemented methods and models (5)-(7) each contain only two of these methods. Model (8) is the proposed model which contains all three methods: multi-headed self-attention, E(N) equivariance, and structural motifs.

\quad In this paper, I employ multi-headed self-attention as Brody et al's GATv2. This layer contains dynamic attention coefficients allowing it to extract complex relationships within the data. To analyze the effectiveness of multi-headed self-attention, I removed the use of attention coefficients during message aggregation and coordinate updates in the EGAT. By comparing models (1) and (4), it is clear that attention mechanisms increase performance due to the improvements in AUROC and accuracy. Furthermore, even in the presence of the other two methods (structural motifs and equivariance), attention mechanisms boost performance as in the increased AUROC and accuracy when comparing model (2) to model (6), model (3) to model (7), and model (5) to model (8). It is believed that the decreases in performance are attributed to the lack of higher-order representations within the data which were previously obtained using these attention coefficients.

\quad To allow the model to break down and understand large anticancer drugs containing hundreds of atoms, I employ structural motifs to extract common reoccurring features. Structural motifs are also shown to improve performance due to the increase in AUROC and accuracy when comparing models (1) and (3), models (4) and (7), and models (2) and (5). It appears that the performance improvements from structural motifs decrease in the presence of multi-headed self-attention and equivariant as there is a minimal increase in AUROC and accuracy when comparing models (6) and (8). The model without the structural motifs cannot effectively encode the drugs due to the large number of atoms in the drug and the limited number of message passaging layers.

\quad The use of equivariant layers also improved the model's performance. Based on the performance differences in models (1) and (2), models (3) and (5), and models (4) and (6), maintaining equivariance improved performance. The use of equivariant layers was nearly as important as attention mechanisms, showing the vast performance improvements by maintaining equivariance. Without the equivariant layers, the model cannot effectively use this positional information, decreasing the model's performance.

\section{Discussions and Conclusions}

In this paper, I propose a novel framework to predict cell line-specific synergistic anticancer drugs. The proposed geometric deep learning framework employs a graph neural network that has multi-headed dynamic attention coefficients and is equivariant to 3D translations, rotations, and reflections. To better represent larger molecules, I also employed structural motifs which extracted common features including ring and non-ring features. 

\quad The proposed method outperformed several SOTA methods on five-fold cross validation experiments including all four synergy metrics and the three dataset types: transductive, unknown combination, and unknown drug. Although the proposed model outperformed SOTA models on all dataset types, it had the greatest performance increases in the unknown combination and unknown drug datasets showing its strong ability to generalize to unseen drugs unlike the other contemporary models. 

\quad In future experiments I would aim to employ different methods to employ equivariance such as spherical harmonics, Clebsch-Gordan coefficients, and Wigner-D matrices. Additionally, I would train the model on larger datasets allowing the model to generalize better. I could also apply this framework in drug-drug interactions as well as antiviral and antifungal synergy. I hope that these methods will be used to expedite the discovery of synergistic anticancer drug combinations and other drug interactions.

\bibliographystyle{IEEEtran}
\bibliography{main}

\end{document}